\documentclass[aps, pra, twocolumn,showpacs,preprintnumbers,amsmath,amssymb]{revtex4}

\newcommand{\ket}[1]{\mbox{$ | #1 \rangle $}}
\newcommand{\bra}[1]{\mbox{$ \langle #1 | $}}
\usepackage[dvips]{graphicx}

\begin{document}

\preprint{}

\title{Quantum broadcast communication}

\author{Jian Wang}

 \email{jwang@nudt.edu.cn}

\affiliation{School of Electronic Science and Engineering,
\\National University of Defense Technology, Changsha, 410073, China }
\author{Quan Zhang}
\affiliation{School of Electronic Science and Engineering,
\\National University of Defense Technology, Changsha, 410073, China }
\author{Chao-jing Tang}
\affiliation{School of Electronic Science and Engineering,
\\National University of Defense Technology, Changsha, 410073, China }


\begin{abstract}
Broadcast encryption allows the sender to securely distribute
his/her secret to a dynamically changing group of users over a
broadcast channel. In this paper, we just consider a simple
broadcast communication task in quantum scenario, which the central
party broadcasts his secret to multi-receiver via quantum channel.
We present three quantum broadcast communication schemes. The first
scheme utilizes entanglement swapping and
Greenberger-Horne-Zeilinger state to realize a task that the central
party broadcasts the secret to a group of receivers who share a
group key with him. In the second scheme, based on dense coding, the
central party broadcasts the secret to multi-receiver who share each
of their authentication key with him. The third scheme is a quantum
broadcast communication scheme with quantum encryption, which the
central party can broadcast the secret to any subset of the legal
receivers.
\end{abstract}

\pacs{03.67.Dd, 03.67.Hk}
\keywords{Quantum key distribution; Quantum teleportation}
\maketitle


%
%
\section{Introduction}
\label{introduction} Quantum cryptography has been one of the most
remarkable applications of quantum mechanics in quantum information
science. Quantum key distribution (QKD), which provides a way of
exchanging a private key with unconditional security, has progressed
rapidly since the first QKD protocol was proposed by Benneett and
Brassard in 1984 \cite{bb84}. A good many of other quantum
communication schemes have also been proposed and pursued, such as
quantum secret sharing
(QSS)\cite{hbb99,kki99,zhang,gg03,zlm05,xldp04}, quantum secure
direct communication (QSDC)
\cite{beige,Bostrom,Deng,denglong,cai1,cai4,jwang1,jwang2,jwang3,hlee,cw1,cw2,tg,zjz}.
QSS is the generalization of classical secret sharing to quantum
scenario and can share both classical and quantum messages among
sharers. Many researches have been carried out in both theoretical
and experimental aspects after the pioneering QSS scheme proposed by
Hillery, Buz\v{e}k and Berthiaume in 1999 \cite{hbb99}. Different
from QKD, QSDC's object is to transmit the secret message directly
without first establishing a key to encrypt it. QSDC can be used in
some special environments which has been shown in Ref.
\cite{Bostrom,Deng}. The works on QSDC have attracted a great deal
of attention. Bostr\"{o}m and Felbinger \cite{Bostrom} proposed a
Ping-Pong protocol which is quasi-secure for secure direct
communication if perfect quantum channel is used. Deng et al.
\cite{Deng,denglong} put forward a two-step QSDC protocol using
Einstein-Podolsky-Rosen (EPR) pairs and a QSDC scheme by using
batches of single photons which serves as quantum one-time pad
cryptosystem. We proposed a multiparty controlled QSDC scheme using
Greenberger-Horne-Zeilinger (GHZ) state and a QSDC scheme based on
the order rearrangement of single photons \cite{jwang2,jwang3}. Lee
et al. \cite{hlee} presented two QSDC protocols with user
authentication. Recently, some multiparty quantum direct
communication schemes which are used to realize the task that many
users send each of their secrets to a central party have been
proposed. Gao et al. \cite{gao1} proposed a QSDC scheme using GHZ
states and entanglement swapping. In their scheme, the secret
messages can be transmitted from two senders to a remote receiver.
They also presented a simultaneous QSDC scheme between the central
party and other $M$ parties using GHZ states, which the $M$ parties
can transmit each of their secret messages to the central party
simultaneously \cite{gao2}. Jin et al. \cite{jin} put forward a
simultaneous QSDC scheme by using GHZ states and dense coding.

Broadcast encryption involves a sender and multi-user (receiver)
\cite{fn94}. The sender first encrypts his content and then
transmits it to a dynamically changing set of users via insecure
broadcasting channels. The broadcast encryption scheme assures that
only privileged receivers can recover the content subscribed and the
unauthorized users cannot learn anything. In this paper, we consider
a simple broadcast encryption task in quantum scenario, called
quantum broadcast communication (QBC), which the central party
broadcasts his secret message to a group of legal receivers via
quantum channel and any illegal receiver cannot obtain the central
party's secret. We then consider a naive scheme that the sender
first establishes a common key with multi-user and then encrypts the
secret with the sharing key. Thus the multi-user can obtain the
sender's secret by decrypting the cypher with the key. In the
present schemes, we try to allow the sender broadcast the secret to
multi-user directly without first establishing a common key to
encrypt it. We then present three QBC schemes based on the ideas in
Ref. \cite{hlee,gao2,jin,zlg01}. In scheme 1, a group of users share
a group key with the central party. After authenticating the users,
the central party broadcasts his secret message to them by using
entanglement swapping \cite{zzhe93}. In scheme 2, each user shares a
authentication key with the central party. The central party first
authenticates the users and then broadcasts the secret to them by
using dense coding \cite{bw92}. In scheme 3, based on quantum
encryption \cite{zlg01}, the central party utilizes controlled-not
(CNOT) operation to encrypt his secret qubit by using the particle
in GHZ state and the designated users also use CNOT operation to
decrypt the central party's secret.

The aim of QBC is to broadcast the secret to legal multi-user
directly. In our schemes, we suppose a trusted third party, Trent,
broadcasts his secret to $r$ legal users, Alice$_1$, Alice$_2$,
$\cdots$, Alice$_r$. Similar to Ref. \cite{hlee}, Trent shares a
secret identity number $ID_i$ ($i=1,2,\cdots,r$) and a secret hash
function $h_i$ ($i=1,2,\cdots,r$) with each user. Only the users'
identities are legal can Trent broadcast his secret to them. Here
the hash function is
\begin{eqnarray}
h: \{0,1\}^l\times\{0,1\}^m\rightarrow\{0,1\}^n,
\end{eqnarray}
where $l$, $m$ and $n$ denote the length of the identity number, the
length of a counter and the length of authentication key,
respectively. Thus the user's authentication key can be expressed as
$AK=h(ID, C)$, where $C$ is the counter of calls on the user's hash
function. When the length of the authentication key is not enough to
satisfy the requirement of cryptographic task. The parties can
increase the counter and then generates a new authentication key. We
denote the authentication keys of Alice$_1$, Alice$_2$, $\cdots$,
Alice$_r$ as $AK_{A_1}=h_{A_1}(ID_{A_1}, C_{A_1})$,
$AK_{A_2}=h_{A_2}(ID_{A_2}, C_{A_2})$, $\cdots$,
$AK_{A_r}=h_{A_r}(ID_{A_r}, C_{A_r})$.

We then give some relations which will be used in our schemes. The
four Bell states and the eight three-particle GHZ states are defined
as
\begin{eqnarray}
\ket{\phi^+}=\frac{1}{\sqrt{2}}(\ket{00}+\ket{11}),\ket{\phi^-}=\frac{1}{\sqrt{2}}(\ket{00}-\ket{11}),\nonumber\\
\ket{\psi^+}=\frac{1}{\sqrt{2}}(\ket{01}+\ket{10}),\ket{\psi^-}=\frac{1}{\sqrt{2}}(\ket{01}-\ket{10}),
\end{eqnarray}
and
\begin{eqnarray}
\label{1}
\ket{\Psi_1}=\frac{1}{\sqrt{2}}(\ket{000}+\ket{111}), \ket{\Psi_2}=\frac{1}{\sqrt{2}}(\ket{000}-\ket{111}),\nonumber\\
\ket{\Psi_3}=\frac{1}{\sqrt{2}}(\ket{100}+\ket{011}), \ket{\Psi_4}=\frac{1}{\sqrt{2}}(\ket{100}-\ket{011}),\nonumber\\
\ket{\Psi_5}=\frac{1}{\sqrt{2}}(\ket{010}+\ket{101}), \ket{\Psi_6}=\frac{1}{\sqrt{2}}(\ket{010}-\ket{101}),\nonumber\\
\ket{\Psi_7}=\frac{1}{\sqrt{2}}(\ket{110}+\ket{001}), \ket{\Psi_8}=\frac{1}{\sqrt{2}}(\ket{110}-\ket{001}),\nonumber\\
\end{eqnarray}
respectively. The four unitary operations which can be encoded as
two bits classical information are
\begin{eqnarray}
I=\ket{0}\bra{0}+\ket{1}\bra{1},\nonumber\\
\sigma_x=\ket{0}\bra{1}+\ket{1}\bra{0},\nonumber\\
i\sigma_y=\ket{0}\bra{1}-\ket{1}\bra{0},\nonumber\\
\sigma_z=\ket{0}\bra{0}-\ket{1}\bra{1}.
\end{eqnarray}
Here the encoding is defined as $I\rightarrow00$,
$\sigma_x\rightarrow01$, $i\sigma_y\rightarrow10$,
$\sigma_z\rightarrow11$. The Hadamard ($H$) operation is
\begin{eqnarray}
H=\frac{1}{\sqrt{2}}(\ket{0}\bra{0}-\ket{1}\bra{1}+\ket{0}\bra{1}+\ket{1}\bra{0}).
\end{eqnarray}

\section{Scheme1: Quantum broadcast communication using entanglement swapping}
In scheme 1, Trent utilizes multi-particle GHZ states and
entanglement swapping to realize quantum broadcast communication,
called QBC-ES. We first present our QBC-ES scheme with two users
(Alice$_1$, Alice$_2$) and then generalize it to the case with many
users (Alice$_1$, Alice$_2$, $\cdots$, Alice$_r$). In the scheme,
Trent broadcasts his secret message to a group of users and all
users have the same authentication key which we call group key (GK).

(S1) Trent prepares an ordered $N$ three-particle GHZ states each of
which is in the state
$\ket{\Psi_1}=\frac{1}{\sqrt{2}}(\ket{000}+\ket{111})_{TA_1A_2}$,
where the subscripts $T$, $A_1$ and $A_2$ represent the three
particles of each GHZ state. Trent takes particle $T$ ($A_1$, $A_2$)
for each state to form an ordered particle sequence, called $T$
($A_1$, $A_2$) sequence. For each GHZ state, Trent performs $I$ or
$H$ operation on particles $A_1$ and $A_2$ according to the group
key, $GK$. If the $i$th value of $GK$ is 0 (1), he performs $I$
($H$) operation on particles $A_1$ and $A_2$. As we have described
in Sec. \ref{introduction}, here $GK=h(ID, C)$. If the length of
$GK$ is not long enough to $N$, new $GK$ can be generated by
increasing the counter until the length of $GK$ is no less than $N$.
Trent also performs randomly one of the two operations \{$I$,
$i\sigma_y$\} on particle $A_1$. He then sends $A_1$ and $A_2$
sequences to Alice$_1$ and Alice$_2$, respectively.

(S2) After receiving $A_1$ and $A_2$ sequences, Alice$_1$ and
Alice$_2$ perform corresponding $I$ or $H$ operations on each of
their particles according to $GK$. For example, if the $i$th value
of $GK$ is 0 (1), Alice$_1$ executes $I$ ($H$) operation on particle
$A_1$. They inform Trent that they have transformed their qubit
using unitary operation according to $GK$. Trent then authenticates
the users and checks eavesdropping during the transmission of $A_1$
and $A_2$ sequences.

(S3) The procedure of authentication and eavesdropping check is as
follows. (a) After hearing from the users, Trent selects randomly a
sufficiently large subset from the ordered $N$ GHZ states. (b) He
measures the sampling particles in $T$ sequence, in a random
measuring basis, $Z$-basis(\ket{0},\ket{1}) or $X$-basis
(\ket{+}=$\frac{1}{\sqrt{2}}(\ket{0}+\ket{1})$,
\ket{-}=$\frac{1}{\sqrt{2}}(\ket{0}-\ket{1})$). (c) Trent announces
publicly the positions of the sampling particles and the measuring
basis for each of the sampling particles. Alice$_1$ (Alice$_2$)
measures the sampling particles in $A_1$ ($A_2$) sequence, in the
same measuring basis as Trent. After measurements, the users
publishes their measurement results. (d) Trent can then authenticate
the users and check the existence of eavesdropper by comparing their
measurement results. If the users are legal and the channel is safe,
their results must be completely correlated. Suppose Trent's
operation performed on particle $A_1$ is $I$. When Trent performs
$Z$-basis measurement on his particle, Alices' result should be
\ket{00} (\ket{11}) if Trent's result is \ket{0} (\ket{1}). On the
contrary, Alices' result should be \ket{++} or \ket{--} (\ket{+-} or
\ket{-+}) if Trent performs $X$-basis measurement on his particle
and gets the result \ket{+} (\ket{-}). Thus if Trent confirms that
the users are legal and there is no eavesdropping, they continue to
execute the next step. Otherwise, he aborts the communication.

(S4) After authenticating the users, Trent announces publicly his
random operations on the particles in $A_1$ sequence and Alice$_1$
performs the same operations on them. Trent divides the remaining
GHZ states into $M$ ordered groups, \{P(1)$_{TA_1A_2}$,
Q(1)$_{T'A_1'A_2'}$\}, \{P(2)$_{TA_1A_2}$, Q(2)$_{T'A_1'A_2'}$\},
$\cdots$, \{P(M)$_{TA_1A_2}$, Q(M)$_{T'A_1'A_2'}$\}, where 1, 2,
$\cdots$, $M$ represent the order of the group and the subscripts
$T$ and $T'$ ($A_1$, $A_1'$ and $A_2$, $A_2'$) denote Trent's
(Alice$_1$'s and Alice$_2$'s ) particles. Trent encodes his secret
on each particle $T$ by using one of the four operations \{$I$,
$\sigma_x$, $i\sigma_y$, $\sigma_z$\}. The parties agree that the
four operations represent two-bit classical message, as we have
described in Sec.\ref{introduction}. Alice$_1$ generates a $M$-bit
random string, $a_1$. For each group, she performs one of the two
unitary operations \{$I$, $\sigma_x$\} on particle $A_1$ according
to $a_1$. For example, if the $i$th value of $a_1$ is 0 (1),
Alice$_1$ executes $I$ ($\sigma_x$) operation on particle $A_1$.
Here Alice$_1$ does not perform any operation on particle $A_1'$.

(S5) Alice$_1$ (Alice$_2$) measures particles $A_1$ and $A_1'$
($A_2$ and $A_2'$) of each group in Bell basis. After measurements,
Alice$_1$ publishes her measurement results, but Alice$_2$ does not
do this directly. According to $GK$, Alice$_2$ transforms her result
by using corresponding unitary operation. If the $(2i-1)$th and
$2i$th values of $GK$ are 00 (01, 10, 11), she performs $I$
($\sigma_x$, $i\sigma_y$, $\sigma_z$) operation on her result and
then publishes the transformed result. If Alice$_2$'s result is
\ket{\phi^+} and the corresponding bits of $GK$ are 01, the
published result by her is \ket{\psi^+}.

(S6) Trent performs Bell basis measurement on particles $T$ and $T'$
of each group and publishes his measurement results. According to
the published information, the users can obtain Trent's secret
message. We then explain it in detail. The state of a group can be
written as
\begin{eqnarray}
\ket{\Psi_1}_{TA_1A_2}\otimes\ket{\Psi_1}_{T'A_1'A_2'}=
\frac{1}{2\sqrt{2}}(\ket{\phi^+_{TT'}}\ket{\phi^+_{A_1A_1'}}\ket{\phi^+_{A_2A_2'}}\nonumber\\
+\ket{\phi^+_{TT'}}\ket{\phi^-_{A_1A_1'}}\ket{\phi^-_{A_2A_2'}}\nonumber\\
+\ket{\phi^-_{TT'}}\ket{\phi^+_{A_1A_1'}}\ket{\phi^-_{A_2A_2'}}+\ket{\phi^-_{TT'}}\ket{\phi^-_{A_1A_1'}}\ket{\phi^+_{A_2A_2'}}\nonumber\\
+\ket{\psi^+_{TT'}}\ket{\psi^+_{A_1A_1'}}\ket{\psi^+_{A_2A_2'}}+\ket{\psi^+_{TT'}}\ket{\psi^-_{A_1A_1'}}\ket{\psi^-_{A_2A_2'}}\nonumber\\
+\ket{\psi^-_{TT'}}\ket{\psi^+_{A_1A_1'}}\ket{\psi^-_{A_2A_2'}}+\ket{\psi^-_{TT'}}\ket{\psi^-_{A_1A_1'}}\ket{\psi^+_{A_2A_2'}}).
\end{eqnarray}
If Trent's encoding operation is $\sigma_x$ which corresponds to
secret bits 01 and Alice$_1$'s random operation is also $\sigma_x$
corresponding to bit 1, \ket{\Psi_1}$_{TA_1A_2}$ is then transformed
to \ket{\Psi_7}$_{TA_1A_2}$ and the state of the group becomes
\begin{eqnarray}
\label{es}
\ket{\Psi_7}_{TA_1A_2}\otimes\ket{\Psi_1}_{T'A_1'A_2'}=
\frac{1}{2\sqrt{2}}(\ket{\psi^+_{TT'}}\ket{\psi^+_{A_1A_1'}}\ket{\phi^+_{A_2A_2'}}\nonumber\\
-\ket{\psi^+_{TT'}}\ket{\psi^-_{A_1A_1'}}\ket{\phi^-_{A_2A_2'}}\nonumber\\
-\ket{\psi^-_{TT'}}\ket{\psi^+_{A_1A_1'}}\ket{\phi^-_{A_2A_2'}}+\ket{\psi^-_{TT'}}\ket{\psi^-_{A_1A_1'}}\ket{\phi^+_{A_2A_2'}}\nonumber\\
+\ket{\phi^+_{TT'}}\ket{\phi^+_{A_1A_1'}}\ket{\psi^+_{A_2A_2'}}-\ket{\phi^+_{TT'}}\ket{\phi^-_{A_1A_1'}}\ket{\psi^-_{A_2A_2'}}\nonumber\\
-\ket{\phi^-_{TT'}}\ket{\phi^+_{A_1A_1'}}\ket{\psi^-_{A_2A_2'}}+\ket{\phi^-_{TT'}}\ket{\phi^-_{A_1A_1'}}\ket{\psi^+_{A_2A_2'}}).
\end{eqnarray}
From the results of Trent and Alice$_1$, Alice$_2$ can deduce
Trent's secret message because the three parties' results correspond
to an exclusive state. For example, the results of Trent and
Alice$_1$ are each \ket{\psi^-_{TT'}} and \ket{\psi^-_{A_1A_1'}} and
Alice$_2$'s original result is \ket{\phi^+_{A_2A_2'}}. According to
Eq. (\ref{es}), the state of the group must be
\ket{\Psi_7}$_{TA_1A_2}\otimes$\ket{\Psi_1}$_{T'A_1'A_2'}$.
Alice$_2$ then knows Trent's secret must be ``01'' because only the
operation $\sigma_x\otimes\sigma_x$ applied on particles $T$ and
$A_1$ can change the state \ket{\Psi_1} into \ket{\Psi_7}. On the
other hand, Alice$_1$ knows $GK$ and she can deduce Alice$_2$'s
original result according to her published result. Similarly, she
can also obtain Trent's secret. Thus Trent broadcasts his secret to
two legal users.

Now, let us discuss the security for the present scheme. The
security requirement for the scheme is that any illegal user cannot
obtain Trent's secret. As long as the procedure of authentication
and eavesdropping check is secure, the whole scheme is secure.
Anyone who has no $GK$ cannot obtain Trent's secret message because
it is impossible to deduce a definite result about the secret from
the published results. We then discuss the security for the
procedure of authentication and eavesdropping check.

At step (S1), Trent performs $I$ or $H$ operation on particles $A_1$
and $A_2$ according to $GK$ which is only shared by the three
parties. If the $i$th bit of $GK$ is 0 or 1, the three-particle GHZ
state becomes
\begin{eqnarray}
\label{security1}
\ket{\Phi_1}&=&\frac{1}{\sqrt{2}}(\ket{000}+\ket{111})=\frac{1}{\sqrt{2}}(\ket{+}\ket{\phi^+}+\ket{-}\ket{\phi^-})\nonumber\\
&=&\frac{1}{2}[\ket{+}(\ket{++}+\ket{--})+\ket{-}(\ket{+-}+\ket{-+})]\nonumber\\
\end{eqnarray}
or
\begin{eqnarray}
\label{security2}
\ket{\Phi_2}&=&\frac{1}{\sqrt{2}}(\ket{0++}+\ket{1--})=\frac{1}{\sqrt{2}}(\ket{+}\ket{\phi^+}+\ket{-}\ket{\psi^+})\nonumber\\
&=&\frac{1}{2}[\ket{+}(\ket{++}+\ket{--})+\ket{-}(\ket{++}-\ket{--})].
\end{eqnarray}
According to Eqs. (\ref{security1}) and (\ref{security2}), if an
eavesdropper, Eve, intercepts particles $A_1$ and $A_2$ and makes a
Bell basis measurement on them, she can obtain partial information
of $GK$. However, Trent performs random $I$ or $i\sigma_y$ operation
on particle $A_1$, which can prevent Eve from eavesdropping the
information of $GK$. As a result of Trent's operation, there are
four possible states \ket{\Phi_1}, \ket{\Phi_2},
\begin{eqnarray}
\label{security3}
\ket{\Phi_3}&=&\frac{1}{\sqrt{2}}(-\ket{010}+\ket{101})=\frac{1}{\sqrt{2}}(\ket{+}\ket{\psi^-}-\ket{-}\ket{\psi^+})\nonumber\\
&=&\frac{1}{2}[\ket{+}(\ket{-+}-\ket{+-})-\ket{-}(\ket{++}-\ket{--})]\nonumber\\
\end{eqnarray}
and
\begin{eqnarray}
\label{security4}
\ket{\Phi_4}&=&\frac{1}{\sqrt{2}}(\ket{0-+}-\ket{1+-})=\frac{1}{\sqrt{2}}(\ket{+}\ket{\psi^-}+\ket{-}\ket{\phi^-})\nonumber\\
&=&\frac{1}{2}[\ket{+}(\ket{-+}-\ket{+-})+\ket{-}(\ket{+-}+\ket{-+})].\nonumber\\
\end{eqnarray}
According to Eqs. (\ref{security1})-(\ref{security4}), Eve cannot
distinguish the four states by using Bell basis measurement. During
the authentication and eavesdropping check, Trent measures his
sampling particles in $Z$-basis or $X$-basis randomly and allows the
users to measure their corresponding particles in the same measuring
basis. Suppose Trent performs $Z$-basis measurement on his particle
and Eve also measures particles $A_1$ and $A_2$ in $Z$-basis. Eve
publishes her measurement result after measurements. If the state is
\ket{\Phi_1} or \ket{\Phi_3}, Eve will not introduce any error
during the process of authentication and eavesdropping check.
However, If the state is \ket{\Phi_2} and \ket{\Phi_4}, Eve will
obtain \ket{00}, \ket{01}, \ket{10} and \ket{11} each with
probability 1/4 and the error rate introduced by her achieves 75\%.
Similarly, when Trent performs $X$-basis measurement and Eve
measures particles $A_1$ and $A_2$ in the same measuring basis, if
the state is \ket{\Phi_2} or \ket{\Phi_4}, Eve will not introduce
any error. But if the state is \ket{\Phi_1} and \ket{\Phi_3}, Eve
will obtains \ket{++}, \ket{+-}, \ket{-+} and \ket{--} each with
probability 1/4 and the error rate is 50\%. Suppose Trent performs
$X$-basis measurement and Eve measures particles $A_1$ and $A_2$ in
Bell basis. When Eve's result is \ket{\phi^+} (\ket{\psi^-}), her
action will not be detected by Trent if she publishes the result
$\ket{++}$ or \ket{--} (\ket{+-} or \ket{-+}). However, when her
result is \ket{\phi^-} (\ket{\psi^+}), if the state is \ket{\Phi_4}
(\ket{\Phi_3}), Trent will detect Eve's eavesdropping. Similarly, if
Trent performs $Z$-basis measurement and Eve executes Bell basis
measurement, Eve's eavesdropping will also be detected by Trent with
a certain probability.

According to Stinespring dilation theorem, Eve's action can be
realized by a unitary operation $\hat{E}$ on a large Hilbert space,
$H_{A_1A_2}\otimes H_{E}$. Then the state of Trent, Alice$_1$,
Alice$_1$ and Eve is
\begin{eqnarray}
\ket{\Phi}=\sum_{T,A_1,A_2\in\{0,1\}}\ket{\varepsilon_{T,A_1,A_2}}\ket{T}\ket{A_1A_2},
\end{eqnarray}
where \ket{\varepsilon} denotes Eve's probe state and \ket{T} and
\ket{A_1A_2} are states shared by Trent and the users. The condition
on the states of Eve's probe is
\begin{eqnarray}
\sum_{T,A_1,A_2\in\{0,1\}}\bra{\varepsilon_{T,A_1,A_2}}\;
\varepsilon_{T,A_1,A_2}\rangle=1.
\end{eqnarray}
As Eve can eavesdrop particle $A_1$ and $A_2$, Eve's action on the
system can be written as
\begin{eqnarray}
\label{security5}
\ket{\Phi}&=&\frac{1}{\sqrt{2}}[\ket{0}(\alpha_1\ket{00}\ket{\varepsilon_{000}}+\beta_1\ket{01}\ket{\varepsilon_{001}}+\gamma_1\ket{10}\ket{\varepsilon_{010}}\nonumber\\
&+&\delta_1\ket{11}\ket{\varepsilon_{011}})+\ket{1}(\delta_2\ket{11}\ket{\varepsilon_{100}}+\gamma_2\ket{10}\ket{\varepsilon_{101}}\nonumber\\
&+&\beta_2\ket{01}\ket{\varepsilon_{110}}+\alpha_2\ket{00}\ket{\varepsilon_{111}}].
\end{eqnarray}
When the state is \ket{\Phi_1}, the error rate introduced by Eve is
$\epsilon=1-|\alpha_1|^2=1-|\delta_2|^2$. Here the complex numbers
$\alpha$, $\beta$, $\gamma$ and $\delta$ must satisfy
$\hat{E}\hat{E}^\dag=I$.

We then generalize the three-party QBC-ES scheme to a multiparty one
(more than three parties). Suppose Trent wants to broadcast his
secret to a group of users, \{Alice$_1$, Alice$_2$, $\cdots$,
Alice$_r$\}. He prepares an ordered $N$ $(r+1)$-particle GHZ states
\begin{eqnarray}
\frac{1}{\sqrt{2}}(\ket{00\cdots0}+\ket{11\cdots1})_{T,A_1,\cdots,A_r}.
\end{eqnarray}
The details of the multiparty QBC-ES is very similar to those of
three-party one. Trent performs $I$ or $H$ operations on particles
$A_1$, $A_2$, $\cdots$, $A_r$ according to GK they shares. He also
performs randomly $I$ or $i\sigma_y$ operation on particles $A_1$,
$A_2$, $\cdots$, $A_{(r-1)}$ and sends $A_1$, $A_2$, $\cdots$, $A_r$
sequences to each Alice$_1$, Alice$_2$, $\cdots$, Alice$_r$. After
receiving the particles, each user performs $I$ or $H$ operations on
their particles according to $GK$. Similar to step (S3), Trent
authenticates the users and checks eavesdropping. If all users are
legal, he announces publicly his operations on particles $A_1$,
$A_2$, $\cdots$, $A_{(r-1)}$ and Alice$_1$, Alice$_2$, $\cdots$,
Alice$_{(r-1)}$ execute the same operations on them. Trent divides
all GHZ states into $N$ ordered groups, [\{P(1)$_{TA_1\cdots A_r}$,
Q(1)$_{T'A_1'\cdots A_r'}$\}, $\cdots$, \{P(N)$_{TA_1\cdots A_r}$,
Q(N)$_{T'A_1'\cdots A_r'}$\}]. He encodes his secret on particles
$T$ using one of the four operations, \{$I$, $\sigma_x$,
$i\sigma_y$, $\sigma_z$\}. Alice$_1$, Alice$_2$, $\cdots$,
Alice$_{(r-1)}$ each perform randomly one of the two operations
\{$I$, $\sigma_x$\} on their particles. Each user measures particles
$A_i$ and $A'_i$ ($i=1,2,\cdots,r$) in Bell basis. After
measurements, Alice$_1$, Alice$_2$, $\cdots$, Alice$_{(r-1)}$
publish their measurement results. Alice$_r$ first transforms her
result according to $GK$ and then publishes the transformed result.
Trent also performs Bell basis measurement on particles $T$ and $T'$
of each group. Thus Trent broadcasts his secret to all legal users,
Alice$_1$, Alice$_2$, $\cdots$, Alice$_(r-1)$ and Alice$_r$. The
security for multiparty QBC-ES scheme is similar to that for
three-party one. As long as the procedure of authentication and
eavesdropping check is secure, the scheme is secure.

Based on entanglement swapping, we can also obtain a multiparty
simultaneous quantum authentication scheme using multi-particle GHZ
state. Here each user shares each of their authentication keys with
Trent. Trent prepares a batch of GHZ states
$\frac{1}{\sqrt{2}}(\ket{00\cdots0}+\ket{11\cdots1})_{t,1,2,\cdots,r}$.
For each GHZ state, he sends particles 1, 2, 3, $\cdots$, $r$ to
Alice$_1$, Alice$_2$, $\cdots$, Alice$_r$, respectively and keeps
particle $t$. Similar to the above method, he divides all GHZ states
into ordered groups and performs randomly one of the two unitary
operations \{ $I$, $i\sigma_y$\} on particle $t$. Each Alice$_i$
($i=1,2,\cdots,r$) performs $I$ or $i\sigma_y$ operation on their
particles according to their authentication keys. Similarly, the
parties perform Bell basis measurements on their particles of each
group. Trent lets the users publish their measurement results and
can then authenticate the $r$ users simultaneously.

\section{Scheme2: Quantum broadcast communication based on dense coding}
In the scheme 1, Trent can only broadcast secret to a group of users
who share a $GK$ with him. Based on dense coding, we present a
quantum broadcast communication scheme, called QBC-DC scheme, which
Trent broadcasts his secret to multi-user \{Alice$_1$, Alice$_2$,
$\cdots$, Alice$_r$\} and the users shares each of their
authentication keys with Trent. We first present a three-party
QBC-DC scheme and then generate it to a multiparty one.

(S1) Trent prepares an ordered $N$ three-particle GHZ states
randomly in one of the eight GHZ states
\{\ket{\Psi_1}$_{TA_1A_2}$,\ket{\Psi_2}$_{TA_1A_2}$,$\cdots$,\ket{\Psi_7}$_{TA_1A_2}$\},
where the subscripts $T$, $A_1$ and $A_2$ represent three particles
of each GHZ state. He takes particle $T$ from each of the GHZ states
to form an ordered particle sequence, called $T$ sequence.
Similarly, the remaining partner particles compose $A_1$ sequence
and $A_2$ sequence. Trent performs one of the two operations \{$I$,
$H$\} on each particle in $A_1$ sequence according to Alice$_1$'s
authentication key, $AK_{A_1}$. Here $AK_{A_1}=h_{A_1}(ID_{A_1},
C_{A_1})$. That is, if the $i$th value of $AK_{A_1}$ is 0 (1), he
performs $I$ ($H$) operation on particle $A_1$. Trent also executes
$I$ or $H$ operation on particle $A_2$ according to Alice$_2$'s
authentication key, $AK_{A_2}$. After doing these, he sends $A_1$
and $A_2$ sequences to Alice$_1$ and Alice$_2$, respectively.

(S2) Alice$_1$ (Alice$_2$) performs corresponding $I$ or $H$
operation on each of her particles according to $AK_{A_1}$
($AK_{A_2}$). For example, if the $i$th value of $AK_{A_1}$ is 0
(1), $I$ ($H$) operation is applied to particle $A_1$. After doing
these, they inform Trent. Trent then authenticates the users and
checks eavesdropping during the transmission of $A_1$ and $A_2$
sequences.

(S3) We then describe the procedure of authentication and
eavesdropping check in detail. (a) Trent selects randomly a
sufficiently large subset from the ordered GHZ states. (b) He
measures the sampling particles in $T$ sequence in $Z$-basis or
$X$-basis randomly. (c) Trent announces publicly the positions of
the sampling particles and the measuring basis for each of the
sampling photons. Alice$_1$ and Alice$_2$ measure their sampling
particles in the same basis as Trent. After measurements, the users
publishes their measurement results. (d) Trent can authenticate the
users and check the existence of eavesdropper by comparing their
measurement results. If the users are legal and the channel is safe,
their results must be completely correlated. For example, the
initial state is \ket{\Psi_1}. Suppose Trent performs $Z$-basis
measurement on particle $T$. If Trent's result is \ket{0} (\ket{1}),
the users' results must be \ket{00} (\ket{11}). If Trent performs
$X$-basis measurement on his particle and gets the result \ket{+}
(\ket{-}), the users' results should be \ket{++} or \ket{--}
(\ket{+-} or \ket{-+}). If any user is illegal, Trent abandons the
communication. Otherwise, they continue to the next step.

(S4) Alice$_1$ and Alice$_2$ each generate a random string, $a_1$
and $a_2$. According to their random strings, they perform one of
the two unitary operations \{$I$, $i\sigma_y$\} on particles $A_1$
and $A_2$, respectively. For example, if the $i$th value of
Alice$_1$'s random string is 0 (1), she performs $I$ ($i\sigma_y$)
operation on particle $A_1$. After their operations, they return
$A_1$ and $A_2$ sequences to Trent.

(S5) Trent selects randomly a sufficiently large subset from the
ordered GHZ states and performs randomly one of the four unitary
operations \{ $I$, $\sigma_x$, $i\sigma_y$, $\sigma_z$\} on each of
the sampling particles in $T$ sequence. He then encodes his secret
message on the remaining particles $T$ by performing one of the four
unitary operations on each of them. Trent measures particles $T$,
$A_1$ and $A_2$ of each GHZ state in three-particle GHZ basis. He
announces publicly the positions of the sampling particles and lets
each user publish their corresponding random operations on the
sampling particles in $A_1$ and $A_2$ sequences. According to his
measurement result, Trent can check the security for the
transmission of the returned particle sequences. When the initial
state is \ket{\Psi_1}, his result is \ket{\Psi_8} and his operation
on particle $T$ is $\sigma_x$, he can deduce Alice$_1$'s and
Alice$_2$'s operations are each $i\sigma_y$ and $I$. He then
compares his conclusion with the operations published by the users
and can decide the security for the transmitting particles. If he
confirms that there is no eavesdropping, he publishes his
measurement results and the initial GHZ states he prepared.
Alice$_1$ and Alice$_2$ can then obtain Trent's secret. For example,
when the initial state is \ket{\Psi_1} and Trent's operation is
$\sigma_x$ (corresponds to his secret 01), $\ket{\Psi_1}$ is
transformed to \ket{\Psi_8}, if Alice$_1$ and Alice$_2$ perform
$i\sigma_y$ and $I$ operations on particles $A_1$ and $A_2$,
respectively. That is,
$\sigma_x\otimes$$i\sigma_y\otimes$$I$\ket{\Psi_1}=\ket{\Psi_8}.
According to her operation performed on particle $A_1$ ($A_2$),
Trent's result \ket{\Psi_8} and the initial state \ket{\Psi_1},
Alice$_1$ (Alice$_2$) can obtains Trent's secret message, 01.

The security for the three-party QBC-DC scheme is based on that for
the procedure of authentication and eavesdropping check. Trent
prepares an ordered $N$ GHZ states each of which is in one of the
eight GHZ states and performs $I$ or $H$ operation on particle $A_1$
and $A_2$ according to each user's authentication key. For each
initial state, there are four possible states after Trent's
operations. For example, the initial state is \ket{\Psi_1}, then the
four possible states are
\begin{eqnarray}
\ket{\Omega_1}&=&\frac{1}{\sqrt{2}}(\ket{000}+\ket{111}),\nonumber\\
\ket{\Omega_2}&=&\frac{1}{\sqrt{2}}(\ket{0+0}+\ket{1-1}),\nonumber\\
&=&\frac{1}{2}[\ket{+}(\ket{\phi^-}+\ket{\psi^+})+\ket{-}(\ket{\phi^+}-\ket{\psi^-})]\nonumber\\
\ket{\Omega_3}&=&\frac{1}{\sqrt{2}}(\ket{00+}+\ket{11-}),\nonumber\\
&=&\frac{1}{2}[\ket{+}(\ket{\phi^-}+\ket{\psi^+})+\ket{-}(\ket{\phi^+}+\ket{\psi^-})]\nonumber\\
\ket{\Omega_4}&=&\frac{1}{\sqrt{2}}(\ket{0++}+\ket{1--}).
\end{eqnarray}
Eve has no information of the user's authentication key. If Eve
intercepts particles $A_1$ and $A_2$ and performs Bell basis
measurement on them, she cannot obtain the information of the user's
authentication key because she cannot distinguish the four states.
During the procedure of authentication and eavesdropping check,
Trent performs randomly $Z$-basis or $X$-basis measurement on his
particle. When Trent performs $Z$-basis measurement and Eve measures
the intercepted particles in the same measuring basis as Trent, if
the state is not \ket{\Omega_1}, it is possible for her to publish a
wrong result after measurements. Similarly, whether Eve utilizes
$Z$-basis, $X$-basis or Bell basis measurement, she will publish a
wrong result with some probability and her action will be detected
by Trent during the procedure of authentication and eavesdropping
check. We can also describe Eve's effect on the system as Eq.
(\ref{security5}). If the initial state is \ket{\Psi_5}, the error
rate introduced by Eve is $\epsilon=1-|\gamma_1|^2$.

We then generalize the three-party QBC-DC scheme to a multiparty
(more than three parties) one which Trent broadcasts his secret to
$r$ users \{Alice$_1$, Alice$_2$, $\cdots$, Alice$_r$\}. He prepares
an ordered $N$ GHZ states each of which is randomly in one of the
$2^{(r+1)}$ $(r+1)$-particle GHZ states
\begin{eqnarray}
\frac{1}{\sqrt{2}}(\ket{ij\cdots k}+\ket{\bar{i}\bar{j}\cdots
\bar{k}})_{T,A_1,\cdots,A_r},
\end{eqnarray}
where $i,j,\cdots,k \in \{0,1\}$ and
$\bar{i},\bar{j},\cdots,\bar{k}$ are the counterparts of
$i,j,\cdots,k$. The details of the multiparty QBC-DC is very similar
to those of three-party one. Trent performs $I$ or $H$ operations on
each particle in $A_1$ ( $A_2$, $\cdots$, $A_r$) sequence according
to $AK_{A_1}$ ($AK_{A_2}$, $\cdots$, $AK_{A_r}$). He then sends
$A_1$, $A_2$, $\cdots$, $A_r$ sequences to Alice$_1$, Alice$_2$,
$\cdots$, Alice$_r$, respectively. After receiving the particle
sequence, each user performs $I$ or $H$ operations on their
particles according to their respective authentication key. Similar
to step (S3) in the three-party QBC-DC scheme, Trent authenticates
the users and checks eavesdropping. If any user is illegal, Trent
aborts communication, otherwise, they continue to the next step.
Alice$_1$, Alice$_2$, $\cdots$, Alice$_r$ perform randomly $I$ or
$i\sigma_y$ operation on their respective particles. After doing
these, they return $A_1$, $A_2$, $\cdots$, $A_r$ sequences to Trent.
Trent first chooses a sufficiently large subset to check the
security for the transmitting particles and performs randomly one of
the four operations on the sampling particles in $T$ sequence. He
also encodes his secret on the remaining particles in $T$ sequence
using the four operations. Trent measures each of $(r+1)$-particle
GHZ states in $(r+1)$-particle GHZ basis. He publishes the positions
of the sampling particles and lets the users announce publicly their
operations on the sampling particles. Trent can then check the
security for the transmission of the returned particle sequences. If
there is no eavesdropping, he publishes his measurement results and
the initial $(r+1)$-particle GHZ states she prepared. Thus Trent
broadcasts his secret to Alice$_1$, Alice$_2$, $\cdots$, Alice$_r$.

\section{Scheme 3: Quantum broadcast communication based on quantum encryption}
In scheme 2, Trent broadcasts his secret to a group of designated
users. We then present a QBC scheme with quantum encryption, called
QBC-QE scheme, which Trent can broadcast his secret to any subset of
the users. The details of the QBC-QE scheme are as follows.

(S1) Trent prepares an ordered $N$ $(r+1)$-particle GHZ states each
of which is in the state
\begin{eqnarray}
\frac{1}{\sqrt{2}}(\ket{00\cdots0}+\ket{11\cdots1})_{T,A_1,\cdots,A_r}.
\end{eqnarray}
The $N$ particles $T$ form $T$ sequence and the $N$ particles $A_i$
($i=1,2,\cdots,r$) form $A_i$ sequence. Trent performs one of the
two operations \{$I$, $H$\} on the particles in $A_i$ sequence
according to Alice$_i$'s authentication key. That is, if Alice$_i$'s
$AK_{A_i}$ is 0 (1), Trent performs $I$ ($H$) on particle $A_i$.
Trent sends $A_1$, $A_2$, $\cdots$, $A_r$ sequences to Alice$_1$,
Alice$_2$, $\cdots$, Alice$_r$, respectively.

(S2) Alice$_i$ ($i=1,2,\cdots,r$) performs corresponding $I$ or $H$
operation on each of her particles according to $AK_{A_i}$. Similar
to the method of step (S3) in QBC-DC scheme, Trent authenticates the
users and checks eavesdropping by performing random $Z$-basis or
$X$-basis measurement. If the users are legal and the channel is
safe, they continue to the next step. Otherwise, Trent stops the
communication.

(S3) Trent utilizes controlled-not (CNOT) operation to encrypt his
secret message. For example, Trent transmits his secret $P=\{p_1,
p_2, \cdots, p_m\}$, where $p_i\in\{0,1\}$ ($i=1,2,\cdots,m$)
represents classical bit 0 or 1, to two users \{Alice$_j$,
Alice$_k$\} ($1\leq j,k\leq r$). He prepares his secret in the state
\ket{p_ip_i}$_{S_jS_k}$, where $S_j$, $S_k$ denote the two particles
of the state. Trent performs CNOT operation on particles $T$, $S_j$
and $S_k$ (particle $T$ is the controller and $S_j$ and $S_k$ are
the targets). Then the GHZ state of the whole quantum system becomes
\begin{eqnarray}
\label{qe1}
\ket{\Upsilon}&=&\frac{1}{\sqrt{2}}(\ket{00\cdots0,p_i,p_i}\nonumber\\
&+&\ket{11\cdots 1,1\oplus p_i,1\oplus
p_i})_{T,A_1,\cdots,A_r,S_j,S_k}.
\end{eqnarray}
According to Alice$_j$'s and Alice$_k$'s authentication keys, Trent
performs corresponding $I$ or $H$ operation on particles $S_j$ and
$S_k$. For example, if the $i$th value of Alice$_j$'s authentication
key is 0 (1), he performs $I$ ($H$) operation on particle $S_j$.
Here we denote the operation performed on $S_j$ ($S_k$) as
$H_{{AK}^i_{A_j}}$ ($H_{{AK}^i_{A_k}}$), where ${AK}^i_{A_j}$
(${AK}^i_{A_k}$) represents the $i$th value of Alice$_j$'s
(Alice$_k$'s) authentication key and $H_0$ ($H_1$) represents $I$
($H$) operation. Thus \ket{\Upsilon} is transformed to
\begin{eqnarray}
\label{qe2}
\ket{\Upsilon'}=\frac{1}{\sqrt{2}}[\ket{00\cdots0,H_{{AK}^i_{A_j}}(p_i),H_{{AK}^i_{A_k}}(p_i)}+\vert 11\cdots1,\nonumber\\
H_{{AK}^i_{A_j}}(1\oplus{p_i}),H_{{AK}^i_{A_k}}(1\oplus{p_i})\rangle]_{T,A_1,\cdots,A_r,S_j,S_k}.\nonumber\\
\end{eqnarray}
Trent then sends $S_j$ and $S_k$ sequences to Alice$_j$ and
Alice$_k$, respectively. To insure the security of the transmission
of $S_j$ and $S_k$ sequences, Trent should insert randomly some
sampling particles into $S_j$ and $S_k$ sequences before sending
them to the users. The aim of inserting the sampling particles is to
make the parties detect Eve's disturbance attack although Eve cannot
obtain any information of Trent's secret message.

(S4) After receiving $S_j$ ($S_k$) sequence, Alice$_j$ (Alice$_k$)
first performs corresponding $I$ or $H$ operation on the
transmitting particles according to $AK_{A_j}$ ($AK_{A_k}$) and then
executes CNOT operation on particles $A_j$ ($A_k$) and $S_j$
($S_k$). For example, if the $i$th value of Alice$_j$'s
authentication key is 1, she performs $H$ operation on the
corresponding particle in $S_j$ sequence and does CNOT operation on
particles $A_j$ and $S_j$ ($A_j$ is the controller and $S_j$ is the
target). According to Eq.(\ref{qe1}) and (\ref{qe2}), Alice$_j$ and
Alice$_k$ can obtain Trent's secret message.

In the above scheme, we just give an example of sending secret to
any two users of $r$ users. Obviously, Trent can send his secret to
any subset of the legal users in the scheme. Strictly speaking, the
present scheme is not a genuine QBC scheme because Trent must
transmit a particle sequence to each user. In view of Trent can
transmit his secret to multi-user directly in this scheme, we still
regard it as QBC. The security for authentication and eavesdropping
check in the scheme is the same as that in QBC-DC scheme. After
confirming the users are legal and insuring the security of the
quantum channel, the GHZ states can be regarded as quantum key.
Trent can then encrypt his secret message using quantum key they
share. If Trent wants to transmit his secret to a subset of the
users, he then performs $I$ or $H$ operation on the encoding
particles according to each designated user's authentication key,
which ensures that only the users in this subset can obtain Trent's
secret. After insuring the security for the transmitting particles,
the designated user decrypts the secret by using quantum key. The
procedure of encryption and decryption in our scheme is the same as
quantum one-time pad, but the quantum key is the GHZ states shared
by the parties. In the scheme, the quantum key can be used
repeatedly for next round of cryptographic task.

\section{summary}
In summary, we have presented three schemes for quantum broadcast
communication. In our schemes, Trent broadcasts his secret message
to multi-user directly and only the legal users can obtain Trent's
secret. In scheme 1, based on the idea in Ref. \cite{gao2}, we
utilizes entanglement swapping to realize a QBC scheme which Trent
sends his secret to a group of users who share a group key with
Trent. In scheme 2, based on the idea in Ref. \cite{jin}, we present
a QBC scheme that Trent broadcasts his secret to multi-user who
share each of their authentication keys with Trent, by using dense
coding. Scheme 3 is based on quantum encryption \cite{zlg01}, which
Trent can broadcast his secret to any subset of the legal users.
Because our schemes utilize block transmission, quantum memory is
necessary. Moreover, compared with classical broadcast encryption
which allows the sender to securely distribute the secret to a
dynamically changing group of users, the present schemes are not
genuine quantum broadcast encryption schemes. That is why we call
them quantum broadcast communication schemes. We hope that our work
will attract more attention and give impetus to further research on
quantum broadcast communication.



\begin{acknowledgments}
This work is supported by the National Natural Science Foundation of
China under Grant No. 60472032.
\end{acknowledgments}

%
%

%
%
\end{document}